\documentclass[11pt]{article}
\usepackage{graphicx}
\usepackage{amssymb}

\advance\textheight by \topskip
\begin{document}
\tolerance=100000
\thispagestyle{empty}
\setcounter{page}{0}

\def\cO#1{{\cal{O}}\left(#1\right)}
\newcommand{\cp}{\hspace{1mm}\mbox{\raisebox{.3mm}{$\diagup$} \hspace{-7.3mm} C \hspace{-3.2mm} P}\hspace{1mm}}
\newcommand{\be}{\begin{equation}}
\newcommand{\ee}{\end{equation}}
\newcommand{\br}{\begin{eqnarray}}
\newcommand{\er}{\end{eqnarray}}
\newcommand{\ba}{\begin{array}}
\newcommand{\ea}{\end{array}}
\newcommand{\bi}{\begin{itemize}}
\newcommand{\ei}{\end{itemize}}
\newcommand{\bn}{\begin{enumerate}}
\newcommand{\en}{\end{enumerate}}
\newcommand{\bc}{\begin{center}}
\newcommand{\ec}{\end{center}}
\newcommand{\ul}{\underline}
\newcommand{\ol}{\overline}
\newcommand{\ra}{\rightarrow}
\newcommand{\sm}{SM}
\newcommand{\as}{\alpha_s}
\newcommand{\aem}{\alpha_{em}}

\def\dis{\displaystyle}
\def\beq{\begin{equation}}
\def\eeq{\end{equation}}
\def\barr{\begin{array}}
\def\earr{\end{array}}

\newcommand{\comment}[1]{}

\def\Ecm{\ifmmode{E_{\mathrm{cm}}}\else{$E_{\mathrm{cm}}$}\fi}
\def\lsim{\buildrel{\scriptscriptstyle <}\over{\scriptscriptstyle\sim}}
\def\gsim{\buildrel{\scriptscriptstyle >}\over{\scriptscriptstyle\sim}}
\def \lum{{\cal L}}

\def\lapp{\mathrel{\rlap{\raise.5ex\hbox{$<$}}
                    {\lower.5ex\hbox{$\sim$}}}}
\def\gapp{\mathrel{\rlap{\raise.5ex\hbox{$>$}}
                    {\lower.5ex\hbox{$\sim$}}}}
\begin{flushright}
{IISc-CHEP/06/07}\\
{LAPTH-1183/2007}\\
\end{flushright}
\vspace{0.3cm}
\begin{center}
{\Large \bf
Top production at the Tevatron/LHC and  nonstandard, strongly interacting spin
one particles}
 \\[0.30cm]
{\large  Debajyoti Choudhury$^{a}$, Rohini M. Godbole${^{b,c}}$, 
Ritesh K. Singh$^{d}$ and Kshitij Wagh$^{e}$}\\[0.35 cm]
$^a$Dept. of Physics and Astrophysics, University of Delhi, Delhi 110 007, India.\\[0.15cm]
$^b$Centre for High Energy Physics, Indian Institute of Science, Bangalore, 560 012, India.\\[0.15cm]
$^c$ Dept. of Theoretical Physics, Tata Institute of Fundamental 
Research, \\Mumbai 400 005, India.\\[0.15cm]
$^d$ Laboratoire d'Annecy-Le-Vieux de Physique Theorique (LAPTH), Chemin de Bellevue,
B.P. 110, F-74941 Annecy-le-Vieux, Cedex, France.\\[0.15cm]
$^e$ Dept. of Physics and Astronomy, Rutgers University, Piscataway, \\
New Jersey 08854-8019, U.S.A.
\end{center}

\vspace{.2cm}

\begin{abstract}
{\noindent\normalsize 

In this note, we consider possible constraints from $t \bar t$ production
on the gauge bosons of theories with an extended strong interaction 
sector such as axigluons
or flavour universal colorons. Such constraints are found to be
competitive with those obtained from the dijet data. The current $t \bar t$ 
data from the Tevatron rule out axigluon masses ($m_A$) up to 910 GeV and 920
GeV at $95\% $  and $90 \% $  confidence levels respectively. For the case 
of the flavour universal colorons, for $\cot \xi = 1$, where $\xi$ is the 
mixing angle, the mass ranges $m_C \lapp 800$ GeV and 
$895 \lapp m_C \lapp 1960$ 
GeV are excluded at $95 \%$ confidence level (C.L.), whereas the same at 
$90 \% $ C.L.  are $m_C \lapp 805$ GeV and $880 \lapp  m_C \lapp  2470$ GeV.
For $\cot \xi =2$ on the other hand,  the excluded  range is 
$m_C \lapp 955 (960)$ GeV and $1030\lapp  m_C \lapp 3250 \
(1020 \lapp m_C \lapp 3250)$ GeV at $95 \% (90 \%)$ C.L. respectively.
We point out that for higher axigluon/coloron masses, even for the dijet 
channel, the limits on the coloron mass, for $\cot \xi = 1$,  may  be 
different than those for the axigluon. We also compute the 
expected forward-backward  asymmetry for the case of the axigluons which
would allow it to be discriminated against the SM as also the colorons. We 
further find that at the LHC, the signal should be visible in the $t \bar t$ 
invariant mass spectrum for a wide range of axigluon and coloron masses 
that are still allowed.  We point out how top polarisation 
may be used to  further discriminate the axigluon and coloron case from the SM 
as well as from each other.
}
\end{abstract}
\newpage

\section{Introduction}
\label{intro}
The Standard Model (\sm) has had unprecedented success in passing
precision tests at the SLC, LEP, HERA and the Tevatron. The agreement
between the directly measured value of the top mass, $m_t$, and the
one indicated by precision measurements~\cite{lepwwg,Alcaraz:2006mx},
has played a crucial role in this test of the \sm\ to the loop level,
providing indirect evidence for the Higgs boson.  However, a direct
verification of the Higgs mechanism of the spontaneous breaking of the
Electroweak (EW) symmetry, thereby providing an understanding of the
mechanism of the mass generation for fermions, is still lacking.  Its
understanding will be one of the focal points of the investigations at
the Large Hadron Collider (LHC) and, thereafter, the International
Linear Collider(ILC)~\cite{myreview}. The top quark, with a mass very
close to the electroweak symmetry breaking scale, is expected to
provide a probe for understanding the phenomenon of symmetry breaking
in the \sm.  For the same reason, any alternative to the Higgs
mechanism almost always involves the top quark~\cite{Hill:2002ap}.
Thus, a study of production and properties of the top quark at the TeV
colliders can be used as a `low' energy probe for any (`high
scale') new physics beyond the \sm, as well as a good probe of
alternates to the Higgs mechanism. An accurate testing of the mass
relations between $m_t$ and $m_W$, as predicted in the \sm, is an
important part of the physics program of any TeV energy
collider~\cite{Weiglein:2004hn}. Clearly, top physics is a very
promising place to look for new physics effects~\cite{Hill:1993hs}.
Already at the Tevatron, this has been a very fruitful area of
investigations~\cite{Cabrera:2006ya,Lannon:2006vm} and one expects a
top factory such as the LHC to be a goldmine for studying the \sm\ as
well as the beyond the \sm (BSM) physics~\cite{Beneke:2000hk}. Recent
discussions in the literature have addressed the possibility of using
$t \bar t$ production at the Tevatron for putting `direct'
constraints~\cite{Guchait:2007ux} on the Kaluza Klein(KK) gluons of
the bulk Randall-Sundrum Model, using their large coupling 
to the $t \bar t$ pairs to probe the same at the LHC through use of
$t$ polarisation, in $t \bar t$ production~\cite{Agashe:2006hk,Lillie:2007yh}, 
as well as using spin-spin correlations in $t \bar t$ production via
KK excitations of the graviton~\cite{Arai:2004yd}. $t \bar t$ pair
production at the Tevatron through a $Z'_t$ in various versions of
Topcolor models has also been discussed~\cite{Harris:1999ya}. The
feasibility of using the $t$ polarisation in $t \bar t$ production
via an EW resonance such as an additional $Z'$ occurring in (say)
unified models or a $Z_H$ which occurs in the Little Higgs
models  has also been recently
explored~\cite{Allanach:2006fy}.  In this note, we revisit the issue of
constraints that the current measurements of $t \bar t$ production at
the Tevatron~\cite{Cabrera:2006ya,Lannon:2006vm} imply for theories
which have a strongly interacting spin-1 particle in addition to the
gluons. We discuss the $t \bar t$ production for the cases
of massive coloured `axigluons' which exist in theories with chiral
colour~\cite{Frampton:1987dn,Frampton:1987ut} and that of the flavour
universal colorons~\cite{Chivukula:1996yr} which exist in certain
versions of extended colour gauge theories. It may be mentioned here
that the case of axigluons and the flavour universal colorons differs
from the other strongly interacting bosons such as the KK gluons or
the ETC theories, in that they do not have preferential coupling to $t
\bar t$. Thus, the sensitivity of $t \bar t$ production to our case
should be expected to differ from that for such models with
enhanced top-couplings. 

 Below, we first summarise the relevant details of the two models as
well as the current limits on the masses of the axigluons and flavour
universal colorons. Then we will show the constraints the current data
on $t \bar t$ production imply for these. We then look at the phenomenology of
the colorons and the axigluons at the LHC as well.

\section{Models.}
In the unifiable chiral colour model~\cite{Frampton:1987dn,Frampton:1987ut}, 
the high energy strong interaction gauge group  $SU(3)_L \times SU(3)_R$  is
spontaneously broken to the usual $SU(3)_{L+R}$ and thus one has an octet
each of massless gluons$(g)$ and massive {\it axigluons} $(A)$ with an axial 
vector coupling, ${\frac {1}{2}} g_s \gamma_\mu \gamma_5 \lambda^a$, where 
$g_s$ is the usual strong coupling and $\lambda_a$ the usual Gell-Mann matrices.
The above mentioned axigluon is common to all the  
different versions of the chiral colour models that exist. 
The strong coupling of $A$ to $q \bar q$  ensures large production 
cross-sections for the axigluons at a hadronic collider, at the same time 
causing the axigluon to have a reasonably large width. 
Considerations of embedding the chiral colour group
into a grand unified theory  implied a `natural' value
for the axigluon mass of the order of the weak scale $\sim 250$ GeV. It was 
soon realised that not only could the axigluon be searched for in the 
dijet channel~\cite{Bagger:1987fz} through the process 
$p \bar p (p) \rightarrow A^* \rightarrow q \bar q$, but also that the 
forward-backward asymmetry caused by the interference between the $g$ and $A$ 
contributions to $q \bar q \rightarrow Q \bar Q$~\cite{Sehgal:1987wi} could 
be used to probe and constrain
the $A$ contribution. The Tevatron searches for new resonances decaying to
dijets~\cite{Abe:1997hm,Giordani:2003ib} exclude the mass range 
$200 < m_A < 1130$ GeV. The lighter axigluon windows were eliminated 
by various considerations such as hadronic decays of the $Z$-boson 
etc.~\cite{Doncheski:1998ny,Yao:2006px}.  $t \bar t$ production
through the process $q \bar q \rightarrow t \bar t$ can be used very
effectively for the axigluon search due to large $q \bar q $ fluxes 
at the Tevatron. The large  top-sample at the Tevatron 
was  shown capable of  providing  reach of the order of 1 TeV for generic new 
physics~\cite{Hill:1993hs}.  The early Tevatron data on top-pair 
production was shown~\cite{Doncheski:1997yj} to disfavour the contribution 
from a lighter axigluon mass window $50 < m_A < 120$ GeV (which was then not 
completely ruled out) at about $1.5 \sigma$ level. The 
analysis can be further sharpened by using the forward-backward  asymmetry 
in the $t \bar t$ production.

The  flavour universal coloron model~\cite{Chivukula:1996yr} 
belongs to a class 
of models of extended colour which arose from the general effort to understand
the mechanism of EW symmetry breaking and the large mass of the top, $m_t$, 
in the
same framework, and wherein a top quark condensate enhances the top 
quark mass and drives the EW symmetry breaking. Specific examples of this
idea are topcolor~\cite{Hill:1991at,Hill:1993hs} and topcolor assisted 
technicolor~\cite{Hill:1994hp}. In general, in the coloron models, it is 
assumed that the colour group at high energies is larger, given by
$SU(3)_I \times SU(3)_{II}$, and breaks at the 
TeV scale to the usual $SU(3)_c$, 
giving rise to an octet of massive, strongly interacting gauge bosons, called
colorons. Variants of the coloron models differ in how the 
different generations of quarks couple to $SU(3)_I$ and $SU(3)_{II}$.  In the 
original model~\cite{Hill:1991at,Hill:1993hs}, the  first two families couple
to $SU(3)_I$ and the third couples to $SU(3)_{II}$. The phenomenology of such
a coloron at the Tevatron and at the LHC, with respect to the $t \bar t$
final states has been discussed~\cite{Hill:1993hs,Dicus:1994sw}. The Universal
flavour coloron model~\cite{Chivukula:1996yr,Simmons:1996fz}, is a variant of 
this idea, 
where all the quarks transform as a $(1,3)$ under this extended gauge group
and the two gauge couplings are $\xi_1,\xi_2$ for $SU(3)_I$ and $SU(3)_{II}$
respectively, with  $\xi_1 \ll \xi_2$. The massive colorons couple to all the
quarks through a ${\frac {1}{2}} g_s \cot \xi  \gamma_\mu \lambda^a$ coupling,
where $\xi$ is the mixing angle given by $\cot \xi = {\xi_2}/{\xi_1}$. 
The mass of the coloron $m_C$ is related to $\xi$,  the strong
coupling $g_s$ and the vacuum expectation value of the scalar $\Phi$ 
which transforms as a $(3,\bar 3)$ under the extended gauge group 
and breaks it down to $SU(3)_c$. 
The coloron, like the axigluon, is also a broad resonance. The model with 
flavour universal coloron can be grafted rather nicely onto the standard one-Higgs-doublet model of EW physics and has a naturally heavy top 
quark~\cite{Popovic:1998vb}.  Electroweak 
precision measurements constrain the $\rho$
parameter and hence the model parameter space, the constraint being given by
$M_c/\cot \xi  \gapp 450 $ GeV~\cite{Chivukula:1995dc}~\footnote{We have 
checked that current precision measurements also imply a similar constraint.}.
Further, the value of $\cot \xi$ is limited from above to $\sim$ 4, by the 
requirement that the model remains in the Higgs phase.

The colorons will 
contribute to the dijet production~\cite{Simmons:1996fz,Bertram:1998wf} 
in almost the same way as the axigluons. 
The  constraints on $m_A$ from the dijet 
data~\cite{Abe:1997hm,Giordani:2003ib}  can also be translated 
into constraints on $m_C$.
However, the coloron contribution will not give rise to any 
forward-backward asymmetry
in $q \bar q$ production as opposed to the case of the axigluons. 
Ref.~\cite{Simmons:1996fz} estimated the possible constraints that may be 
obtained with  the $b$--tagged dijets, using the analysis constraining the 
topgluon production and decay, accounting for the difference in the 
possible decay channels and the width in the two cases. Further, it was 
claimed~\cite{Simmons:1996fz} that while using the approximation of incoherent 
sum of the background and the signal, the signal strengths in the dijet channel
for the flavour universal coloron for $\cot \xi =1$ 
will be the same as that of an axigluon of the same mass. It was further
claimed~\cite{Simmons:1996fz}, that since the expected cross-sections for 
the coloron, in the above approximation, increase with $\cot \xi$, the 
constraints implied by the dijet analysis for the axigluons also give the most 
conservative constraint for the coloron.  At present, the best quoted bounds
for axigluon and colorons come from the Tevatron dijet data which rule out 
masses up to~$\simeq 980$  GeV~\cite{Giordani:2003ib,Yao:2006px}. As we point 
out later, for heavier coloron masses, some of the above statements need
to be amended.

\section{Axigluon and 
Coloron contribution to $t \bar t$ production at hadronic 
colliders.}
\label{Formulae}

At the tree level, 
the presence of an axigluon $A$ or the coloron $C$ can 
affect $t \bar t$ production only as far as the $q\bar q$-initiated 
subprocess 
is concerned, leaving the $gg$-initiated subprocess unaltered. 
Since the $q \bar q$ fluxes are dominant over the 
$gg$ fluxes at the Tevatron, the possible presence of a axigluon/coloron 
resonance (the $s$-channel $q \bar q \rightarrow A^* (C^*) \rightarrow 
t \bar t$ diagram is the only new contribution) can affect
the total rate of $t \bar t$ production significantly. 
This can be easily understood by realising that, at the Tevatron, even in the 
absence of these exotic bosons (i.e., even for the \sm\ alone), the 
contribution to the total $t \bar t$ cross-section from the $q \bar q$ 
initial state dominates over the $g g$ contribution by a factor of 
$~\sim {\cal O} (10)$, depending on the PDF's, choice of scale etc.

In the presence of an axigluon, the parton-level differential cross section 
for the $q \bar q$-initiated process is modified and  for final state 
$t (\bar t)$ with helicity $\lambda (\bar \lambda)$ is given by, 
\beq
\barr{rcl}
\dis \frac{d \sigma}{d \hat t} (q \bar q \to t(\lambda)  \bar t(\bar \lambda))
& = & \dis 
\frac{\pi \, \alpha_s^2}{9 \, \hat s^2} 
      \; \Bigg\{
       \Bigg[ (1 + \lambda \, \bar \lambda ) \;
       \frac{4 \, m_t^2}{\hat s} \, 
                ( 1 -  c_\theta^2 )
       + (1 - \lambda \, \bar \lambda ) \; ( 1 + c_\theta^2 ) 
\Bigg]
\\[3ex]
& & \dis \hspace*{3em} +
\frac{(1 -  \lambda \, \bar \lambda) }
     {(\hat s - m_A^2)^2 + \Gamma_A^2 \; m_A^2} \; 
\left[
\hat s^2 \, \beta^2 \, ( 1 + c_\theta^2 ) 
+
4 \, \hat s \, (\hat s - m_A^2)\; \beta \, c_\theta  \, 
\right]
\Bigg\} \ ,
\earr
\label{eq:axiprodn_pol}
\eeq
where $\hat s $ is the invariant mass of the $t \bar t $ system 
with $\beta (\equiv \sqrt{1 - 4 \, m_t^2 / \hat s})$ and $\theta$ 
being the top-velocity and the scattering angle in the parton 
centre of mass frame respectively; $\lambda$  and $\bar \lambda $  which 
are the helicities (as distinct from chirality) of the top and the anti-top,
and take values $\pm 1$. 
Note that there are no terms linear in the helicities, as would have been 
present if  the intermediate boson (axigluon) were to have both 
vectorial and axial couplings. The width 
$\Gamma_A$ is given by 
\beq \dis
\Gamma_A \equiv \sum_{q} \Gamma (A \to q \bar q) 
\approx \frac{\alpha_s \, m_A}{6} \; \left[5 + 
\left( 1 - \, \frac{4 \, m_t^2}{m_A^2} \right)^{3/2} \right] \ .
\label{eq:axiwidth}
\eeq
Summing over the top polarizations, the expression for the cross-section for
$t \bar t$ production for the $q \bar q$ initial state becomes,
\beq
\barr{rcl}
\dis \frac{d \sigma}{d \hat t} (q \bar q \to t \bar t)
& = & \dis 
\frac{2 \, \pi \, \alpha_s^2}{9 \, \hat s^2} 
      \; \Bigg\{
       \Bigg[ 
       \frac{4 \, m_t^2}{\hat s} \, 
                ( 1 -  c_\theta^2 )
       + ( 1 + c_\theta^2 ) 
\Bigg]
\\[3ex]
& & \dis \hspace*{2em} +
\frac{\hat s^2 \, \beta^2 \, ( 1 + c_\theta^2 ) + 4 \, \hat s \, (\hat s - m_A^2)\; \beta \, c_\theta  }
     {(\hat s - m_A^2)^2 + \Gamma_A^2 \; m_A^2} \; 
\Bigg\} \ ,
\earr
\label{eq:axiprodn_unpol}
\eeq
and reduces to Eq.~[2.2] of ~\cite{Bagger:1987fz} in the limit of zero
quark (top) mass. Note the existence of the term odd in $c_\theta$ here, which
is due to the interference between the gluon and the axigluon amplitude. We may
mention here that our Eq.~\ref{eq:axiprodn_unpol} does not 
agree with Eq.[2] of Ref.~\cite{Sehgal:1987wi}. The interference term in
their Eq. [2]  is proportional to $\beta^2$ instead of $\beta$ as in our 
Eq.~\ref{eq:axiprodn_unpol}; further as per their Eq.[2] the square of the 
axigluon exchange amplitude is proportional to the gluon exchange amplitude,
apart from the obviously different propagator, which can not be
true for massive quarks in the final state.  
Needless to mention of course that
the first term in eq.~\ref{eq:axiprodn_unpol}, corresponding to the \sm, agrees
with the standard QCD expression~\cite{Combridge:1978kx}, now available in 
textbooks on the subject. Since the 
contribution of the 
$gg$ initial state~\cite{Combridge:1978kx} (to be added incoherently)
remains unchanged we refrain from reproducing the 
formulae here.

In the presence of the coloron, the differential cross-section  for production
of production of $t (\bar t)$ with helicities $\lambda (\bar \lambda)$ 
respectively,  reads for the $q \bar q$ initial state: 
\beq
\barr{rcl}
\dis \frac{d \sigma}{d \hat t} (q \bar q \to t(\lambda)  \bar t(\bar \lambda))
& = & \dis 
\frac{\pi \, \alpha_s^2}{9 \, \hat s^2} 
      \; \Bigg\{
 (1 + \lambda \, \bar \lambda ) \;
       \frac{4 \, m_t^2}{\hat s} \, 
                ( 1 -  c_\theta^2 )
       + (1 - \lambda \, \bar \lambda ) \; ( 1 + c_\theta^2 ) 
\Bigg\}
\\[2ex]
& & \dis \hspace*{3em}
\left| 1 + \frac{\hat s \; \cot^2 \xi}
                {\hat s - m_C^2 + i \, \Gamma_C \, m_C} \right|^2 \ .
\earr
\label{eq:colprodn_pol}
\eeq
The width is now given by
\beq \dis 
\Gamma_C \equiv \sum_{q} \Gamma (C \to q \bar q) 
= 
   \frac{\alpha_s \; \cot^2 \xi }{6 } \, m_C\; 
   \left\{ 5 + 
        \left[1  + 2 \, \frac{m_t^2}{ m_C^2} \right] \; 
\left( 1 - \, \frac{4\, m_t^2}{m_C^2}\right)^{1/2} 
\right\} \ .
\label{eq:colwidth}
\eeq
Once again, summing over the polarization gives us for the differential 
cross-section
\beq
\barr{rcl}
\dis \frac{d \sigma}{d \hat t} (q \bar q \to t \bar t)
& = & \dis 
\frac{2 \, \pi \, \alpha_s^2}{9 \, \hat s^2} 
      \; \Bigg\{
       \frac{4 \, m_t^2}{\hat s} \, 
                ( 1 -  c_\theta^2 )
       + ( 1 + c_\theta^2 ) 
\Bigg\}
\; 
\left| 1 + \frac{\hat s \; \cot^2 \xi}
                {\hat s - m_C^2 + i \, \Gamma_C \, m_C} \right|^2 \ .
\earr
\label{eq:colprodn_unpol}
\eeq
As far as the coloron is concerned, the net effect of the addition of a simple
vectorial interaction is to just change the propagator from $1/\hat s$ to
$1/\hat s + \cot^2 \xi / (\hat s - m_C^2 + i m_C \Gamma_C
)$~\cite{Hill:1993hs}, in the $q \bar q \rightarrow t \bar t$ amplitude.  
Our Eq.~\ref{eq:colprodn_unpol} agrees with Eq.[3.7]
of Ref.~\cite{Simmons:1996fz} and also with Eq.[3.3],
the corresponding expression for dijet cross-section,
when the limit of zero top mass is taken.  In the absence of the 
coloron contribution the term in Eq.~\ref{eq:colprodn_unpol} proportional
to $\cot \xi$ is absent and then the expression trivially reduces to the
usual QCD contribution for the $q \bar q$ initial state~\cite{Combridge:1978kx}.
Again the presence of colorons does not modify the contribution of the $gg$ 
initial state from its \sm\ form. 

\begin{figure}[!htb]
\vspace*{-0.8in}
\begin{center}
\includegraphics*[scale=0.55]{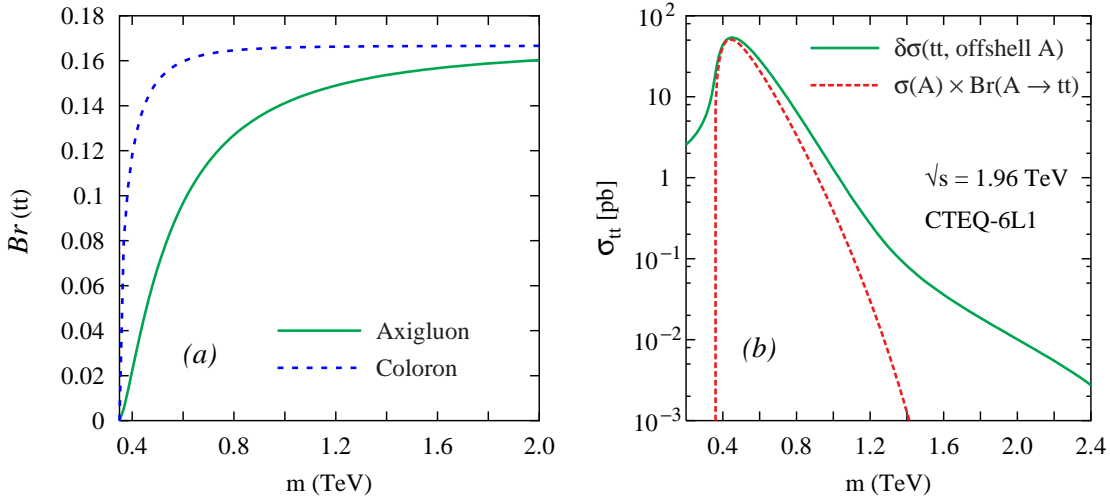}
\end{center}
\vspace*{-1.2in}
\caption{\em (a) The branching fraction of axigluon/coloron to a
$t \bar t$ pair as a function of the boson mass;
 (b) A comparison of the deviation of the total $t \bar t$ cross section
caused by the presence of an axigluon (solid line) with the
resonant production followed by decay.
}
\label{fig:branching}
\end{figure}

We now comment on certain differences between the two cases (axigluons
vs. colorons) as also on the (non-)applicability of certain
approximations made in the literature in the context of dijet
production. For one, a comparison of Eqs.(~\ref{eq:axiwidth} \&
~\ref{eq:colwidth}), shows that the two decay widths differ for 
$m_A \, (m_C) \gapp 2 \, m_t$ (a range not the focus of discussions in
Refs.~\cite{Bagger:1987fz,Simmons:1996fz}) even for $\cot \xi = 1$.
This becomes even more pronounced when we consider the branching
fraction into a $t \bar t$ pair (see Fig.\ref{fig:branching}$a$). 
Thus  the branching fraction into dijets will also be different for a 
coloron with $\cot \xi =1$ from an axigluon of the same mass,
for boson masses $\gapp 2 m_t$, particularly  in the range $0.5\ {\rm TeV}\ 
< m_{\rm boson}\ < 1.5$ TeV, the relative difference in the
branching fractions  however being smaller than that for the $t \bar t$ case. 
Furthermore, the large widths immediately
imply that the (narrow-width) approximation of resonant production and
subsequent decay is no longer a good one. This is borne out by
Fig.\ref{fig:branching}$b$, wherein we compare the $t \bar t$
cross-section at the Tevatron expected, for an axigluon, in the
resonant (incoherent) approximation with the exact value of the
deviation in $t\bar t$ cross section obtained by integrating over the
(large) width of the resonance. To be more specific, we compare 
$\sigma (A) \times BR (A \rightarrow t \bar t)$
with $\delta \sigma \equiv \sigma_A (t \bar t) - \sigma_{SM} (t \bar t)$.
One sees that the approximation could
underestimate the cross-section by up to a factor of 3-4, even for
axigluon  masses as low as $\sim 1$ TeV, which are well within the
reach of the Tevatron. For the case of the coloron, the situation is
even more complicated as we discuss below.

The differential cross-sections for the two cases
differ markedly in the interference term between the gluon-exchange
and the nonstandard strong gauge boson (axigluon or coloron)-exchange 
amplitudes.  For the axigluon, this contribution, being a parity-odd one, 
does not contribute to the total cross section, but 
gives rise to a forward-backward asymmetry. For the coloron, on the other 
hand, the interference term does indeed contribute to the 
total cross section and thus is of great importance when integrating 
over the non-resonant contributions. 
Note further that the interference terms between the coloron
and gluon-exchange contributions changes sign as the subprocess centre
of mass energy $\sqrt{\hat s}$ passes through $m_C$, and, depending on the
value of $\cot \xi$, can even reduce  the total integrated cross-section below
the \sm\ value.  Thus, the 
interference term has different behaviour for the coloron and the axigluon.
As can be also seen from Eqs.~\ref{eq:axiprodn_unpol} and \ref{eq:colprodn_unpol}, even the term 
in the cross-section proportional to the square of the propagator
of the unstable strong boson (axigluon or coloron) is different in the two 
cases for massive quarks in the final state.  
 
The observations above have several  implications.  First, it is clear that 
for $m_A, m_C  \lapp  1.5 $ TeV, as far as
the $t \bar t$ production cross-section is concerned, (even on resonance) 
the expectations for a coloron  with $\cot \xi =1$ will be different from 
that for an axigluon of the same mass, unlike the case of dijet cross-sections
where only massless quarks are involved. Even for the latter case, for the
heavier colorons  now being looked for in the Tevatron dijet data, for 
$\cot \xi =1$, the $\sigma \times B$ , is no longer the same as that of an 
axigluon of the same mass. As was already mentioned, an equality between 
these,  claimed in Ref.~\cite{Simmons:1996fz} and used in all the 
analyses~\cite{Giordani:2003ib,Yao:2006px} so far, is true only for 
relatively low values of the boson masses. It may be pointed that even in 
that case the (approximate) equality is true only when the resonant 
contribution dominates and one considers
only the incoherent sum of background and the signal. With increasing mass
of the boson, the width increases, necessitating the inclusion of the 
interference term to the total cross-section and the approximation of
incoherent sum of the background and the signal is no longer valid. For 
the case of the
colorons, the effect of the parity-even interference term gives 
a non monotonic dependence of the total integrated cross-section on $m_C$ and 
$\cot \xi$. Thus it means that for heavier colorons, a value of mass $m_C$
excluded at $\cot \xi =1$  need not be excluded at higher values of $\cot \xi$.

\section{Numerical Results and constraints from $t \bar t$ production at Tevatron}

Using the formulae presented in section~\ref{Formulae} (and including 
the contribution 
from the $gg$ initial state~\cite{Combridge:1978kx}), we now proceed to
calculate the $t \bar t$ cross-section at the Tevatron Run II 
($\sqrt{s} = 1.96$) and assess the implications of the current data. 
For all our computations, we use the CTEQ-6L1 parton 
distributions~\cite{Pumplin:2002vw}, with a choice of $Q^2 = m_t^2$ for the
factorization scale and $m_t= 175$ GeV. For the SM process, corresponding 
$K$--factor corresponding to this choice of PDF and the scale, 
amounts to $1.08$~\cite{Campbell:2006wx}.  
As mentioned earlier, at the 
Tevatron, the \sm\ production process is dominated by the $q \bar q$ 
initial states, and since both the color and tensorial structure of 
the process under consideration  is very similar to the SM process, it is 
expected that the use of the SM $K$-factor for the cross-section including the 
effects of the axigluon/coloron contribution is well justified. This 
is what we shall use henceforth. We have verified that except very close 
to the peaks, the cross-sections are stable with respect to a change in 
the scale of the hard process (and/or the choice of the parton distributions) 
as long as the corresponding correct $K$-factor~\cite{Campbell:2006wx} is used.
While the smallness of $K (1.08)$ for our choice of parton density, is a 
reflection of the rather large value of $\alpha_s(M_Z) = 0.130$
used in CTEQ-6L1, note that this choice of parton densities is not a 
special one. 
\begin{figure}[!h]
\begin{center}
\vspace*{-1.0cm}
\includegraphics[width= 10 cm, height= 10 cm]{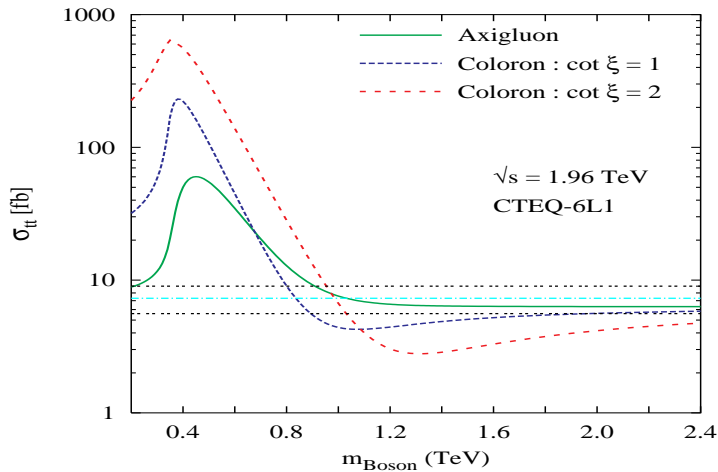}
\vspace*{-1.8cm}
\caption{\em $t \bar t$ production cross-section at the Tevatron as a function 
of the axigluon (coloron mass) and constraints on it from the current 
$t \bar t$ production data. The solid (green) line 
corresponds to the axigluon case. The short- (blue) and long-dashed (red) 
lines correspond to the flavour universal coloron for $\cot \xi = 1$ and 
$\cot \xi=2$ respectively. The horizontal lines correspond
to the current central value from the CDF experiment~\cite{Cabrera:2006ya}
and the $95 \% $  confidence level bands. 
We have used CTEQ-6L1 parton distribution functions evaluated at $Q = m_t$ 
and included the appropriate $K$--factor~\protect\cite{Campbell:2006wx}.}
\label{fig:cs_tev}
\end{center}
\end{figure}

In Fig.\ref{fig:cs_tev}, the solid (green) line shows our 
predictions for the $t \bar t$ production cross-section as a 
function of the axigluon mass. This may be compared with the current 
experimental data which gives (CDF Run II results averaged over 
all channels)~\cite{Cabrera:2006ya}
\be
\sigma(p + \bar p \to t + \bar t + X; \sqrt{s} = 1.96 \, {\rm TeV})
 = 7.3 \pm 0.5 \, {(stat)} \pm 0.6 \, {(syst)} \pm 0.4 \, {(lum)} \ {\rm pb}\ .
\label{refdata}
\ee
In Fig.\ref{fig:cs_tev}, the central value is denoted by the horizontal 
grey dashed line, whereas the  sidebands (dashed black and magenta lines 
respectively) give the $95 \%$ --Confidence Level (C.L.) limits (obtained 
by adding the errors in quadrature). As can be seen from the plot, the data 
rule out axigluon masses($m_A$) up to 910 GeV at  $95 \%$ C.L. 
It can be easily seen that these  are comparable to the current 
constraints~\cite{Giordani:2003ib,Yao:2006px} available from the 
Tevatron dijet data. 
As already mentioned before, predictions for the $t \bar t$ production  
for the coloron (even for $\cot \xi =1$) differ from that for the axigluon 
of the same mass.  In particular, due to the destructive (parity-even) 
interference for $\hat s < m_C^2$, the cross-section
shows a dip as a function of $m_C$ and rises again. That the 
extent of this interference depends crucially
on the magnitude of $\cot \xi$ can be understood from 
the discussion following Eq.(\ref{eq:colprodn_unpol}).
As can be clearly seen, in this case, the data rule out values of $m_C$  
below $800$ GeV at $95 \% $ confidence level and also the 
mass range between $895$ GeV to $1960$ GeV at $95 \%$ confidence level, for 
$\cot \xi = 1$.  At $90\%$ C.L. the exclusion is for $m_C < 805$ GeV
and for masses  between $880$ GeV to $2470$ GeV. For $\cot \xi =2$ on the 
other hand, the excluded  range at $95 \%$ C.L. is $m_C \lapp 955 $ GeV and 
$1030 
\lapp m_C \lapp 3250$ GeV. At $99.99 \%$ C.L., the same is, e.g., 
$m_C \lapp 930 $ GeV and $1110 \lapp m_C \lapp 1860$ GeV.

\begin{figure}[tbh]
\begin{center}
 \vspace{-1cm}
\includegraphics*[width= 12 cm, height= 10 cm]{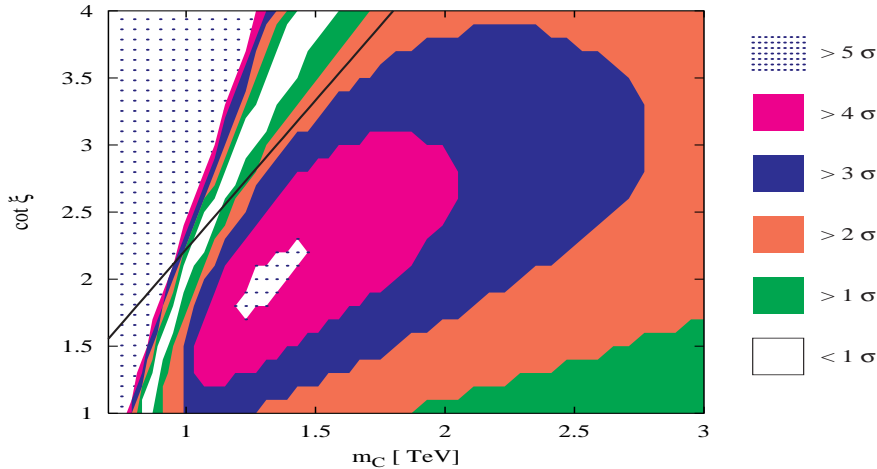}
 \vspace{-1.5cm}
\caption{\em Exclusion region in the $\cot \xi$ -- $m_C$ plane using the
$t \bar t$ data at Tevatron~\protect\cite{Cabrera:2006ya}. 
The solid curve shows the constraint imposed by the 
$\rho$ parameter $m_C/\cot \xi \gapp 450$. We restrict the plot to 
$\cot \xi < 4.0$ for reasons mentioned in the text.}
\label{fig:exclusion}
\end{center}
\end{figure}

It should be noted that the fact of the consistency of certain domains 
of model parameters with the data cannot be interpreted as evidence for 
the existence of colorons, simply because the data are also consistent 
with the SM predictions as well. Rather, the Tevatron data
may be used to rule out regions in the $m_C$--$\cot \xi$ plane
and we display the corresponding results in Fig.\ref{fig:exclusion}.
Once again, we are witness to the r\^ole of the aforementioned interference 
effects. The thin white sliver is consistent with the Tevatron data at the 
$1 \sigma$ level. Thus the values of $\cot \xi$--$m_C$ lying in this
sliver are consistent with the current data. 
(Note that any quantitative statement depends rather critically
on the exact value of $m_t$ used, due to the strong dependence of the
cross-section on $m_t$ near the edge of the phase space boundary, as the 
case is at the Tevatron.)  As one goes out of the region of the sliver, we find
that the regions of the $\cot \xi$--$m_C$ plane are excluded as they 
predict cross-sections either above or below the upper and lower bound on the
experimentally measured $t \bar t$ cross-section of Eq.~\ref{refdata} at a 
given level of confidence. Thus excluded regions on either side of the
thin sliver, are excluded either from above (the cross-section being  too high) 
or from below (the cross-section being too low).
As one moves away to the left of this sliver, the cross section increases
sharply, leading to the very thin bands as depicted in Fig.\ref{fig:exclusion}.
Moving to the right though, the cross section initially falls, and then 
rises again, both changes being slower than that to the left of the 
$1\sigma$ sliver. The different rate of change of the cross-section
with $m_C$ to the left and right of the sliver is a reflection of the behaviour 
already seen in Fig.\ref{fig:cs_tev}; it can be traced back to the interplay 
of the interference effect---as in Eq.(\ref{eq:colprodn_unpol})---and the 
parton density distributions. More concretely, this results in the iso-cross 
section contours being closed curves 
roughly focused at $\sim$(1.2 TeV, 1.7). Indeed, the bands to the left of the 
$1\sigma$ sliver, and the sliver itself, are but parts of larger 
closed bands, not quite apparent in the figure as
we have restricted the range of 
$\cot \xi$  such that one always stays within the Higgs phase of the 
model~\cite{Simmons:1996fz}. The figure is also reflective of the fact that,
for a given $m_C$, the cross-section could be a non-monotonic function of 
$\cot \xi$. For small $m_C$, the cross section does increase monotonically 
with 
$\cot\xi$, but for larger values of $m_C$, the cross-section first 
{\em falls} as we increase $\cot\xi$ from unity (a reflection of the 
destructive interference operative for $\hat s < m_C^2$) and then increases 
once the coloron amplitude overwhelms the gluon amplitude. 
Such effects, for example, result in the island of $5 \sigma$ exclusion 
approximately centred at $\sim$(1.2 TeV, 1.7).
Also shown in  Fig.\ref{fig:exclusion}
is the limit imposed by the 
$\rho $ parameter, namely $m_C/\cot \xi \gapp 450$ GeV; we see that 
even below this
line there is a patch of white ($1\sigma$ agreement).

Apart from the above mentioned differences between the 
axigluon and coloron in the total cross-section, the two cases are also 
distinguished by the parity odd and even nature of the interference term, as
was already mentioned.  This gives rise to a 
forward-backward (FB) asymmetry~\cite{Sehgal:1987wi} 
at the Tevatron for the
production of dijets as well as heavy quarks in the final state. For the
latter, the FB asymmetry can be measured using the decay leptons from the heavy
quark, in the present case the $t$ quark. The numerical value for this
asymmetry had been calculated in Ref.~\cite{Doncheski:1997yj} using the 
formulae of Ref.~\cite{Sehgal:1987wi}, which are incorrect. 
\begin{figure}[tbh]
\begin{center}
\vspace*{-1.0cm}
\includegraphics[width= 10 cm, height= 10 cm]{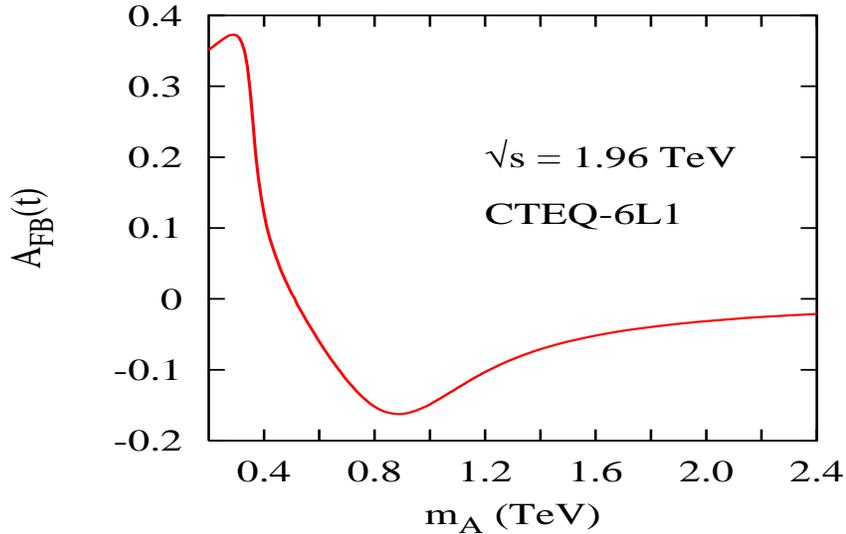}
\vspace*{-1.8cm}
\caption{\em Forward backward asymmetry expected in the 
$t \bar t$ production due to axigluon contribution as a function of
the axigluon mass $m_A$.}
\label{fig:FB}
\end{center}
\end{figure}
We present, in Fig.\ref{fig:FB}, the FB asymmetry expected at the Tevatron, 
calculated using our Eq.~\ref{eq:axiprodn_unpol}, as a function of $m_A$.
Note again that the $gg$ initiated, FB-symmetric contribution has been
included in the denominator as it should be.  The asymmetry is sizable and 
may be used to obtain further constraints.  
As for the SM background to such an asymmetry, it is easy to see that 
the {\it small} parity violating EW contribution does not interfere with
the strong amplitude, due to the different color structure and hence
contribution to the FB asymmetry from the $Z$ exchange can be safely
neglected. An additional contribution to the same 
also accrues~\cite{Kuhn:1998jr,Kuhn:1998kw}, the dominant part coming 
from the absorptive part of correction to the Born-level 
amplitude for $q \bar q \to t \bar t$. 
In fact, a similar contribution to the FB charge asymmetry exists
even in the context of pure 
QED~\cite{Berends:1973fd} in $e^+ e^- \rightarrow \mu^+ \mu^-$. 
The computations of the leading contributions to this  asymmetry of 
Refs.~\cite{Kuhn:1998jr,Kuhn:1998kw}, find that an asymmetry as large 
as $4$--$5 \% $ is possible within the SM. This raises the issue of the 
separability of asymmetry caused by the axigluon contribution.
Within the SM, the asymmetry arises from the contributions of a specific 
subclass of real and virtual radiative corrections and hence it has a 
particular phase space distribution.  As Figures 5 and 9 of 
Refs.~\cite{Kuhn:1998jr,Kuhn:1998kw} show, the asymmetry coming from the QCD 
source is larger at lower values of sub-process center-of-mass energy and at 
larger production angles of the top quark.  On the other hand, the 
contribution to the FB asymmetry coming from a resonance will have a 
distinctive $m_{t \bar t}$  dependence and thus the two may be separated 
by making appropriate cuts on these two quantities. Such a study, though 
interesting, is beyond the scope of the present investigation. 
In view of the use of FB asymemtry in searching for  unusual $t \bar t$ 
resonances, an analysis of this asymmetry taking into account higher order
effects  is certainly worth doing, but again quite beyond the scope of the 
current work.

\section{$t \bar t$ production at the LHC due to coloron/axigluons}
\begin{figure}[tbh]
\begin{center}
\vspace*{-1.0cm}
\includegraphics[width= 10 cm, height= 10 cm]{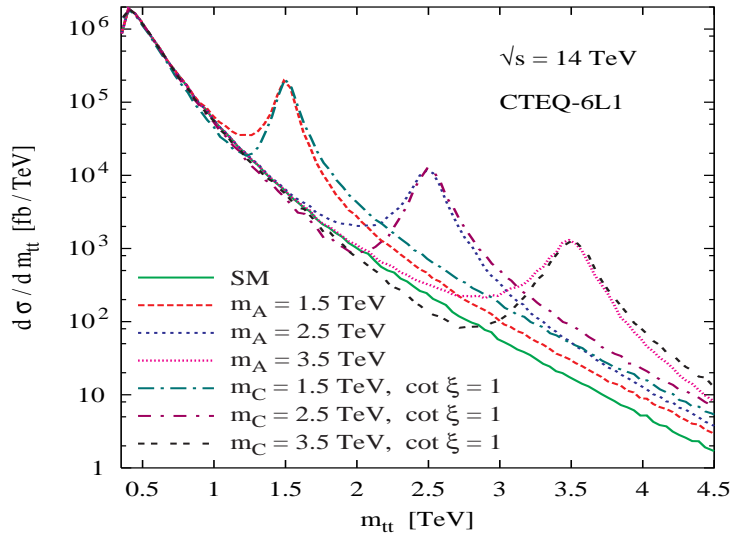}
\vspace*{-1.8cm}
\caption{\em The expected 
$m_{t \bar t}$ 
spectrum at the LHC in presence of either axigluons or colorons of 
a specified mass. Also shown, for comparison, are the SM expectations.
}
\label{fig:mtt_lhc}
\end{center}
\end{figure}

The situation of course is very different at the LHC as the $q \bar q$ 
fluxes are substantially smaller than the $gg$ fluxes and $t \bar t$ 
production is dominated by contribution from  $gg$ initial state. However,
this dominance is less severe as we go to larger $t \bar t$ invariant masses.
Since, at the LHC, we would typically be interested 
in exploring larger axigluon (coloron) masses, it is wiser to concentrate 
on a data sample that has enhanced sensitivity to the mass range in question.
Hence, in this case, instead of looking at the total integrated
$t \bar t$ cross-sections, we consider the effect of the axigluon/coloron
contribution on the $m_{t \bar t}$ spectrum. The large size of the top
sample expected at the LHC ($\sim$ 8 million pairs for 10 $fb^{-1}$ 
integrated luminosity), makes a study of such differential distribution 
meaningful.  In Fig.~\ref{fig:mtt_lhc},
we show the $d \sigma/d m_{t \bar t}$ distribution as a function of 
$m_{t \bar t}$ at the LHC for both the cases, viz. the axigluon and 
the coloron. Let us reiterate that  contributions from both, the $q \bar q$
and $g g$,  initial states have been included in the numbers presented in 
Fig.~\ref{fig:mtt_lhc}. 
As can be seen  from these plots, the resonances, though very broad, will show
up clearly over the continuum.  Note, for example,
for the first peak in Fig.~\ref{fig:mtt_lhc}, even assuming a $10 \%$ 
efficiency,  there will be about $\sim 10^4$ events at $10$ fb$^{-1}$. This 
should allow a cross-section measurement to a $\%$ level.
It may also be noticed that, precisely at the resonance, 
the cross-sections for the axigluon and the coloron for $\cot \xi =1$ for the
same mass are indeed the same. (For, at that point, the 
additional cross section is well-approximated by 
$\sigma(p \bar p \to A/C) \times Br (A/C \to t \bar t)$, and for such 
large values of the boson masses, the branching fractions are nearly 
identical.)
The distributions also show evidence of the aforementioned 
interference effects in case of the coloron. While this is destructive 
for $m_{t \bar t} < m_C$, it is the opposite for $m_{t \bar t} > m_C$. 
It can be easily seen that, for the same mass, the coloron cross-section should
be asymptotically bigger than the axigluon cross-section by about
a factor 2 as is seen in Fig.~\ref{fig:mtt_lhc}. 

The LHC being a $pp$ machine, the possibility of 
constructing a FB asymmetry does 
not exist.
Since the axigluon has only a pure pseudo-vectorial coupling, a measurement 
of the net $t$ polarisation will also not probe the difference in the 
nature of the coupling between the axigluon and the 
coloron
unlike the case of
(say) an extra $Z'$ present in many extensions of the Standard 
Model~\cite{Allanach:2006fy}.
Indeed as Eqs.(\ref{eq:axiprodn_pol}\&\ref{eq:colprodn_pol}) show, the 
only dependence on polarizations is through the product of the 
two ($t$ and $\bar t$) helicities and thus only a variable 
sensitive to this product can exploit this aspect. An example is afforded by 
\beq \dis
{\cal R}_\Delta(m_{tt}) \equiv 
    \left[ \int_{m_{tt} - \Delta}^{m_{tt} + \Delta} \, d m_{tt} \, 
                    \frac{d \sigma_{-}}{d \, m_{tt}} \right]
  \;    \left[ \int_{m_{tt} - \Delta}^{m_{tt} + \Delta} \, d m_{tt} \, 
                    \frac{d \sigma_{+}}{d \, m_{tt}}  \right]^{-1}
\ ,
\label{eq:defn_R}
\eeq
where $\sigma_{\pm}$ refer to the cross sections for the 
product of the $t$ and $\bar t$ helicities to be $\pm 1$ respectively. 
Ideally, the interval $\Delta$ is to be chosen so as to maximize the 
sensitivity, and would nominally be a function of the width of the 
boson in question and the experimental accuracy in measuring $m_{tt}$. 
Rather than do this, we adopt two nominal values of 
$\Delta = 0.1 \, m_{\rm Boson}$ and $\Delta = 0.2 \, m_{\rm Boson}$, given 
the fact that the first choice well approximates 
$\Delta \simeq \Gamma_{\rm Boson}$. 

\begin{figure}[tbh]
\begin{center}
\vspace*{-1.0cm}
\includegraphics[width= 15 cm, height= 10 cm]{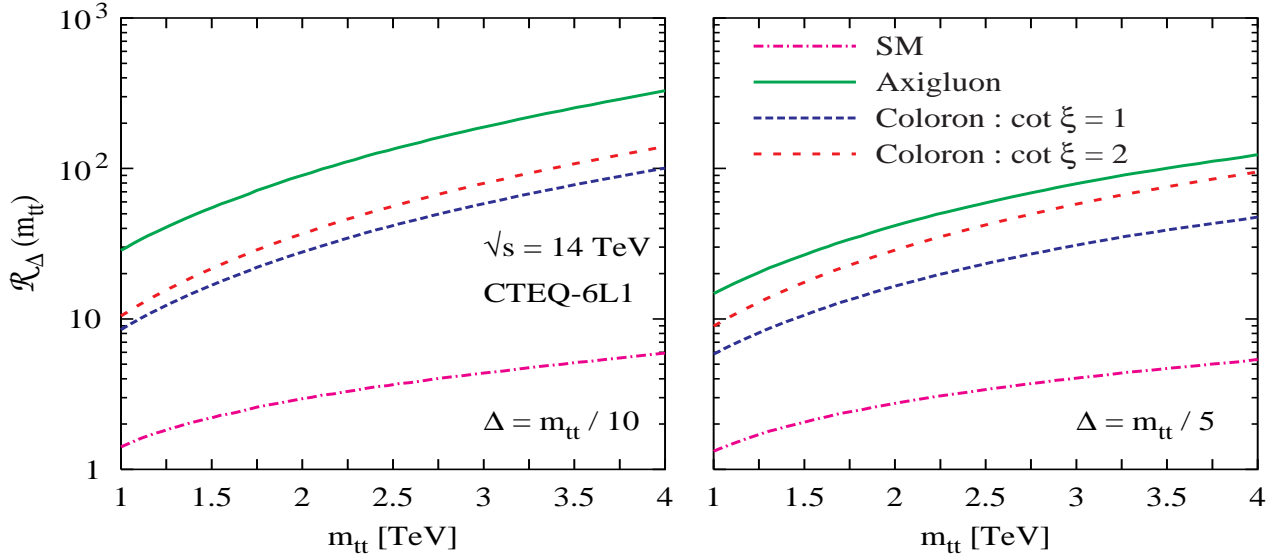}
\vspace*{-1.8cm}
\caption{\em The ratio of the partial cross-sections 
${\cal R}_\Delta(m_{tt} = m_{\rm Boson})$---
see Eqn.(\ref{eq:defn_R})--- as a function of the boson mass. The 
two panels correspond to different values of $\Delta$. 
}
\label{fig:lr_lhc}
\end{center}
\end{figure}

In Fig.\ref{fig:lr_lhc}, we exhibit this ratio, evaluated at 
$m_{tt} = m_A \,(m_C)$ as a function of the boson mass. Clearly, in 
either case, the enhancement, with respect to the SM,
of the $\lambda \bar \lambda = - 1$ channel is much more
marked than that for the $\lambda \bar \lambda = + 1$ channel.  Furthermore, 
this also differentiates between the axigluon and coloron modes for 
comparable values of couplings. This discriminator is in addition over the 
invariant mass distribution (as displayed in Fig.\ref{fig:mtt_lhc}). 
While it is true that an experimental measurement of this 
ratio (essentially a measurement of the correlation of the spin of the
$t$ quark with the spin of the $\bar t$) 
would be a challenging one, it will not be impossible. 
In fact, possibilities of probing the
heavy KK graviton resonances or the Heavy Higgs through such spin-spin 
correlations, have been studied in detail
~\cite{Smolek:2004uf,Hubaut:2005er} in the context of the ATLAS detector. 
This study shows that the asymmetry 
between like spin and unlike spin-pair can be measured with a precision
of $4 \%$, with $10$ fb$^{-1}$ luminosity.  It is also
worth pointing out that the dependence on the integral interval $\Delta$ while 
substantial, is not so large as to invalidate the efficacy of this observable.

\section{Summary}
We have investigated the contribution to the $t \bar t$ production at the 
Tevatron, in the presence of strongly interacting, non-standard spin-1 
particles such as axigluons or flavour universal colorons.  We find that the
CDF Run-II  data on $t \bar t$  production rule out axigluon masses 
$m_A$ up to $910 (920)$ GeV at $95\%$--$(90\%)$ confidence level. For the 
flavour universal colorons, the constraints depend on the mixing angle 
$\xi$ in a non monotonic way. For  $\cot \xi = 1$, the data rule out 
$m_C \lapp 800 (805)$ GeV at $95\% (90 \%)$ confidence level, as well as 
$895 \lapp  m_C \lapp  1965~(880 \lapp m_C \lapp 2475)$ GeV at $95\% (90 \%)$ 
C.L.  On the other hand, for $\cot \xi =2$, the same data rule out 
$m_C \lapp 955 (960)$ GeV and $1030 \lapp m_C \lapp 3200 \; (1020 \lapp m_C 
\lapp 3250) $ GeV at  $95\%$--$(90 \%)$ C.L. 
We correct the formulae in literature for the axigluon mediated 
$t \bar t$ production
cross-section as well as for the FB asymmetry expected at a $\bar p p$ collider
in this case.  We also point out that for the range of large axigluon/coloron
masses, being probed at the Tevatron, the approximation of the incoherent sum
of the resonance and the background, used normally in the analyses so far,
will  not be sufficient due to the large width of the boson. This fact, along 
with the differences in the $t \bar t$ decay width between a coloron and an 
axigluon, makes the expected cross-section for a flavour universal coloron for 
$\cot \xi =1$, different from that for an axigluon of the same mass,
even for the dijet channel.  We have further shown that
the former  also has the effect of making the contribution of the parity 
even interference term negative for the coloron case, causing the 
cross-section to dip below the SM value for certain values of $m_C$ depending 
on $\cot \xi$. We compute the FB asymmetry using the corrected formulae and 
show that it can be used to discriminate the axigluon contribution from the SM 
as well as from the 
coloron case. We further find that at the LHC, in spite of the dominance of 
the $gg$ initial state, the axigluon/coloron contribution can be seen
clearly in the $m_{t \bar t}$ spectrum. We further suggest a variable 
constructed using top/anti-top polarizations which can discriminate the
axigluon or coloron contribution from that of a gluon as well as
from each other.

\section*{Acknowledgments}
We wish to acknowledge discussions with Prof.  S.D. Rindani.
D.C. acknowledges support from the Department of Science and Technology, 
India under project number SR/S2/RFHEP-05/2006. 
R.M.G., R.K.S and K.W.  wish to acknowledge support from Indo French 
Centre for Promotion of Advanced Research Project 3004-B. 


\end{document}